\documentclass[12pt,preprint]{aastex}

\begin{document}

\title {MID-INFRARED SPECTRA OF DUST DEBRIS AROUND MAIN-SEQUENCE STARS\footnote{Based on observations with the NASA {\it Spitzer Space Telescope}, which is operated by the California Institute of Technology for NASA}}

\author{M. Jura\footnote{Department of Physics and Astronomy, University of California, Los Angeles CA 90095-1562}$\,\,$, C. H. Chen\footnote{Jet Propulsion Laboratory, California Institute of Technology, Pasadena CA 91109}$\,\,$, E. Furlan\footnote{Department of Astronomy, Cornell University, Ithaca NY 14853-6801}$\,\,$, J. Green$^{5}$, B. Sargent$^{5}\,$, W. J. Forrest\footnote{Department of Physics and Astronomy, University of Rochester, Rochester NY 14627-0171}$\;$,  D. M. Watson$^{5}$,  D. J. Barry$^{4}$, P. Hall$^{4}$, T.L. Herter$^{4}$, J. R. Houck$^{4}$, G. C. Sloan$^{4}$, K. Uchida$^{4}$, P. D'Alessio\footnote{Centro de Radioastronomia y Astrofisica, UNAM,  58089 Morelia, Michoacan, Mexico  }$\;$, B. R. Brandl\footnote{Sterrewacht Leiden, Leiden 2300 RA Netherlands}$\;$,  L. D. Keller\footnote{Department of Physics, Ithaca College, Ithaca NY 14850}$\;$, F. Kemper$^{2,}\,$\footnote{{\it Spitzer} Fellow}$\,$, P. Morris\footnote{{\it Spitzer} Science Center, California Institute of Technology, Pasadena CA 91125}$\,\,\,\,$, J. Najita\footnote{National Optical Astronomy Observatories, Tucson AZ 85726-6732}$\,\,\,\,$, N. Calvet\footnote{Harvard Smithsonian Center for Astrophysics, Cambridge MA 02138}$\,\,\,$, L. Hartmann$^{12}$, P. C. Myers$^{12}$}

\begin{abstract}
 
We report  spectra obtained with the {\it Spitzer Space Telescope} in the ${\lambda}$ = 14 -- 35 ${\mu}$m range of 19 nearby main-sequence stars with infrared excesses.      The  six stars with strong dust emission show no recognizable spectral features, suggesting that the bulk of the emitting particles have diameters larger than
10 ${\mu}$m.  If the observed dust results from
 collisional grinding of larger solids, we infer minimum  masses of the parent body population  between 0.004 M$_{\oplus}$ and 0.06 M$_{\oplus}$.  We estimate grain production rates of ${\sim}$10$^{10}$ g s$^{-1}$ around   ${\lambda}$ Boo and HR 1570; selective accretion of this matter may help explain their peculiar surface abundances.  There appear to be inner truncations 
in the dust clouds at 48 AU, 11 AU, 52 AU and 54 AU around  HR 333, HR 506,  HR 1082 and  HR 3927, respectively.

\end{abstract}

\section{INTRODUCTION}
The discovery with IRAS of a large infrared excess around Vega (Aumann et al. 1984) initiated the
detailed study of other planetary systems. We can now use the dust
emission from main sequence stars to constrain models for the origin and evolution of these environments (Lagrange, Backman \& Artymowicz 2000, Zuckerman 2001) with the long-term goal of developing a more comprehensive understanding of the formation and evolution of planets and
related minor bodies. 
Most previous studies of the dust around  main-sequence stars have been  restricted to using broad-band infrared fluxes (e. g. Habing et al. 2001, Spangler et al. 2001).  Here, we report measurements with the Infrared Spectrograph\footnote{The IRS was a collaborative venture between Cornell 
University and Ball Aerospace Corporation funded by NASA 
through the Jet Propulsion Laboratory and the Ames Research 
Center.} (IRS, Houck et al. 2004) on the {\it Spitzer Space Telescope} (Werner et al. 2004) of 19 main-sequence
stars.  

\section{OBSERVATIONS}

 We obtained spectra of main sequence stars with previously reported IRAS excesses 
 with both the Short-Low (5.2 -- 14.5 ${\mu}$m) and Long-Low  (14.0 -- 38.0 ${\mu}$m; ${\lambda}/{\Delta}{\lambda}$ ${\sim}$ 90) modules on the IRS.     Table 1 lists the six targets in our
sample with strong infrared excesses.  We did not
detect a substantial excess around 13 other stars. An example of a star without a strong excess is
HR 4732; its spectrum (from {\it Spitzer} aor 0003579392) is shown in Figure 1. 
The other stars without
strong excesses are  HD 16157 (aor 0003554560), HD 221354 (aor 0003565842), HR 818 (aor 0003554816),  HR 1338 (aor 0003555584), HR 2015 (aor 0003556608), HR 3220 (aor 0003557888), HR 3862 (aor 0003558400), HR 5447 (aor 0003560192), HR 8085 (aor 0003564544), HR 8549 (aor 0003588608), HR 8799 (aor 0003565568) and Ross 128 (aor 0003559168).     

In order to avoid time-consuming peak-up on our relatively bright targets with accurately known positions, we operated
the observatory in  IRS Spectral Mapping mode where a 2x3 raster
(spatial x dispersion) centered on the star is performed. The slit
positions were separated by half of the slit width in the dispersion direction,
and by a third of the slit length in
the spatial direction.

We carried out the bulk of the reduction and analysis of our spectra with the IRS team`s
SMART program (Higdon et al 2004). We started with intermediate products from the {\it Spitzer}
IRS data-reduction pipeline, that lacked only stray light and flat-fielding corrections.
We extracted point-source spectra for each nod position
using a uniformly weighted column profile for which the variable width scales with the width of the
instrumental pointspread function. We also made similar spectral extractions for observations
of our photometric standard star, $\alpha$ Lac (A1V). Then we divided
each target spectrum, nod position by nod position, by the spectrum of  the standard
star, multiplied the quotient by the appropriate template spectrum (Cohen et al 2003), and
averaged the resulting spectra from corresponding nod
positions. We plot the resulting spectra of our targets with IR excesses in Figure 1.
Based upon comparisons of IRS spectra of non-variable calibration sources to ground-based, IRAS, and {\it Spitzer} IRAC flux densities, we estimate
that the absolute spectrophotometric accuracy of these spectra is 10\% (1$\sigma$).  It is important to
distinguish this absolute photometric accuracy, which describes the vertical positions of
the spectra on our plots relative to the flux density scale, from the point-to-point
fluctuation in the spectra, which is a measure of the precision of our
spectral feature strengths and therefore of our ability to identify spectral features and
compare them to our models.  In general, the accuracy of our spectral feature identifications
and feature
strengths is limited by the point-to-point scatter (currently much larger than the system noise,
in bright objects) that is visible in the spectra. Therefore almost all features that
significantly exceed the fluctuations, and are wider than the two-pixel IRS spectral
resolution element, are significant. The lack of spectral features that we refer to
in this paper is based on this criterion.

\section{INTERPRETATION AND ANALYSIS}

 To interpret the data for the infrared excess, we need 
luminosities, $L_{*}$, and 
ages, $t_{*}$, of the stars.  We take  distances, $D_{*}$, from 
{\it Hipparcos}, apparent magnitudes from the Yale Bright Star Catalog, assume no interstellar reddening, and use the bolometric corrections given by Flower (1996) to infer the total luminosities.  For four stars, we adopt published estimated ages.  For HR 506, the age is highly uncertain
since Zuckerman \& Song (2004) and Decin et al. (2000, 2003) estimate 0.3 Gyr and 3.0 Gyr, respectively. In view of the recently measured lithium abundance
of 2.6 on the usual 12-point logarithmic scale (Israelian et al. 2004), we adopt the younger age.     For  HR 333 and HR 3927, we assume  ages   of 0.1 Gyr, representative of dusty main
sequence A-type stars (Song et al. 2001).
 
\subsection{Grain Size}
The dust
continua for ${\lambda}$ $>$ 20 ${\mu}$m show no spectral features.      If the particles are spheres of diameter $a$, then any spectral features that might be present are
weakened if $a$ $>$ ${\lambda}/{\pi}$ (e. g. Spitzer 1978, Wolf \& Hillenbrand 2003).   Our data thus suggest that the diameter of the grains is larger than 10 ${\mu}$m, as predicted by  models such as those of Krivov, Mann \& Krivova (2000) and found for the particles that produce the zodiacal light in the Solar System (Fixsen \& Dwek 2002).

\subsection{Spatial Distribution and Dynamics}    

A standard  model of dust debris systems
assumes that collisions among a population of large, unseen parent bodies  produces smaller, detectable  dust particles (Lagrange et al. 2000, Zuckerman 2001). Some of the best studied systems such as HR 4796A exhibit a well defined
ring of material (Schneider et al. 1999).   Very approximately,  
for circumstellar envelopes with $L_{30}/L_{*}$ $>$ 10$^{-4}$ (where $L_{30}$ is the characteristic luminosity  of the infrared excess in the 30 ${\mu}$m spectral window, or ${\nu}L_{\nu}$ at 30 ${\mu}$m), particle-particle collisions
typically are the most important grain destruction  mechanism (e. g. Krivov, Mann \& Krivova 2000, Wyatt \& Dent 2001).  
In these relatively dusty environments, the grains do not migrate  far  before
they  are fragmented into sufficiently small
pieces that they are driven out of the system by radiation pressure.  Thus, 
the spatial distribution of the dust reproduces the spatial distribution
of the parent bodies which might be determined by the history of planet
formation in the system.    
 In contrast, in less dusty 
envelopes with $L_{30}/L_{*}$ $<$ 10$^{-4}$, mutual collisions are not so important, and the grains spiral inwards under the Poynting-Robertson drag. In such models, the spatial distribution of the 
grains need not reproduce  that of the parent bodies.  

  Motivated by these models of dust debris, we consider two possible fits to the reduced data.  First, we fit the spectra by a function of the form:
\begin{equation}
F_{\nu}\;=\;K_{1}\,{\nu}^{2}\;+\;K_{2}{\nu}^{-1}
\end{equation}
where $K_{1}$ and $K_{2}$ are constants.  
The first term in equation (1) describes the photospheric emission while the second term represents 
the infrared excess.   We show in Figure 1 the comparison of our best fits   with the data, and we 
list in Table 1 our values for $K_{2}$ and $L_{30}/L_{*}$ where $L_{30}$  is $4{\pi}D_{*}^{2}K_{2}$.  
 Equation (1) fits the spectra for the two stars with the lowest values of $L_{30}/L_{*}$, HR 1570 and ${\lambda}$ Boo.  However, 
 the infrared excesses of four stars require another model since fits using equation (1) either underestimate the observed photospheric fluxes
measured at ${\lambda}$ ${\leq}$ 12 ${\mu}$m (HR 333 and HR 3927) or the measured IRAS and/or ISO fluxes (Decin et al. 2000)
at ${\lambda}$ = 60 ${\mu}$m (HR 506 and HR 1082). 

Our second method  to model the infrared excess is to fit a single temperature
to the dust particles so the total flux is fit with a function of the form:
\begin{equation}
F_{\nu}\;=\;C_{1}\,{\nu}^{2}\;+\;C_{2}\,B_{\nu}(T_{ex})
\end{equation}  
where $B_{\nu}$ is the Planck function.  In principle,
 $C_{1}$ must equal $K_{1}$, but the absolute level of our spectra is somewhat uncertain because of imprecise calibration. 
As can be seen in Figure 1, we can fit our infrared spectra adequately with equation (2) with the values of $T_{ex}$ given in Table 1 ranging from 72 K
to 114 K. 

The spectral energy distribution of the infrared excess can  test the models.
If Poynting-Robertson drag dominates, the spectrum varies as ${\nu}^{-1}$ (e. g. Buitrago \&
Mediavilla 1985), the dust is smeared over a wide range of radii and equation (1) should fit the data.  In contrast, in a model where collisions limit the lifetime of the
particles, then inward drift is not as important and the dust is more concentrated near its birthplace.  In environments with $L_{30}/L_{*}$ $>$ 10$^{-4}$,  there might be a dust ring or at least a well defined inner hole to the dust distribution which achieves a characteristic maximum temperature. Within the limited IRS spectral range,  equation (2) can fit the data.  Inner dust truncations might be
the result of the formation of low mass companions which might be planets (Kenyon \& Bromley 2002), but other models can produce rings as well (Takeuchi \& Artymowicz 2002).    Also, since $T_{ex}$ is near 100 K, if water ice is an important constituent of the material,
there may be substantial grain destruction by sublimation which contributes
to the observed deficiency of warm grains (Jura et al. 1998).

We now estimate some parameters for the circumstellar
dust.  We assume that the particles
radiate like black bodies so that particles at temperature $T$ lie at distance $D$ from the star:
\begin{equation}
D\;=\;\left(\frac{L_{*}}{16{\pi}T^{4}{\sigma}_{SB}}\right)^{1/2}.
\end{equation}
For the two stars where the spectra can be fit by equation (1), there is no
constraint on the maximum temperature and thus no evidence for any inner
truncation in the dust distribution.  
There are four stars  where  we use equation (2) to fit the data, and we
thus find a characteristic 
 grain temperature, which, from equation (3), we can translate
 into a distance from the star.  
Using the values of $T_{ex}$ and $L_{*}$ given in Table 1, we find values for the inner boundary of the dust of  48 AU, 11 AU, 52 AU and 54 AU around  HR 333, HR 506,  HR 1082 and  HR 3927, respectively.
 Consistent with models for those dust debris systems 
 where frequent collisions lead to smaller particles which are then ejected by radiation pressure, the stars with the strongest evidence for inner holes in the dust distribution  have the highest values of $L_{30}/L_{*}$.    If the inner holes are  caused by inner planets, then in the future, with sufficiently high angular resolution, it may be possible to identify azimuthal asymmetries
in the dust which might indicate  gravitational perturbations by
this unseen system of planets (e. g. Wyatt 2003).

We can also use our data to make a rough estimate of the minimum mass of the parent bodies, $M_{PB}$.  
 Since Poynting-Robertson drag provides a maximum grain lifetime,  then in an approximate steady state, we can write (Chen \& Jura 2001):
\begin{equation}
M_{PB}\;{\geq}\frac{4L_{30}t_{*}}{c^{2}}
\end{equation}
This estimate of $M_{PB}$  is insensitive to the grain size and composition; 
  the minimum values of $M_{PB}$ derived from equation (4) are shown in Table 1. In systems with $L_{30}/L_{*}$ $>>$ 10$^{-4}$, collisions among dust grains occur often, the Poynting-Robertson lifetime is thus much longer than the
collision time, and we may severely underestimate $M_{PB}$.  There may also
be large objects such as planets that are not experiencing rapid collisions and which do not contribute to the detected dust.

\section{DISCUSSION}
   We show in Table 1 that the inferred lower bounds to  $M_{PB}$ 
range from 0.004 M$_{\oplus}$ to 0.06 M$_{\oplus}$. The summed mass of the KBOs in the Solar System  is uncertain and is currently estimated  to be between 0.02 M$_{\oplus}$ and 0.1 M$_{\oplus}$ (Bernstein et al. 2003; Luu \& Jewitt 2002).    Typical analogs of the Kuiper Belt may have total masses ${\leq}$ 
 0.1 M$_{\oplus}$ (Jura 2004).  Since we observed stars with notable infrared excesses, we probably selected objects with particularly massive systems of parent bodies.  In the Solar System, the value of the infrared luminosity of the dust is ${\sim}$ 10$^{-7}$ L$_{\odot}$ (e. g. Fixsen \& Dwek 2002).
 Either the mass of parent bodies around our sample of stars is much
larger than the inferred minimum, or the collision rate of these objects
is substantially greater than in the Kuiper Belt of the Solar System.

Two of the stars in Table 1, ${\lambda}$ Boo itself  and HR 1570,  are ${\lambda}$ Boo stars, a subclass  of  ${\sim}$2\% of all main-sequence A-type stars that is defined as having nearly solar abundances of C, N and O, but marked deficiencies of Fe elements by factors of 10--100 (Paunzen et al. 2002).
The explanation for this abundance pattern is not known.  
One theory to account for their
surface abundances is that they have preferentially accreted some circumstellar matter (Venn \& Lambert 1990), and within the star's atmosphere there is an element-dependent chemical separation controlled by  gravitational
settling and radiative acceleration at the base of the thin outer convection zone of the star (Turcotte \& Charbonneau 1993).  
As shown in Jura (2004), the rate at which circumstellar dust is being produced scales
as the infrared excess luminosity such that ${\dot M}$ ${\sim}$ $L_{30}/c^{2}$  
where $c$ is the speed of light.  We therefore find from the results in Table 1 that ${\dot M}$  ${\sim}$10$^{10}$ g s$^{-1}$ for both ${\lambda}$ Boo and HR 1570.            
According to Turcotte \& Charbonneau (1993),  hydrogen-rich gas infall rates of  ${\geq}$10$^{12}$ g s$^{-1}$ with selective accretion of ${\geq}$10$^{10}$ g s$^{-1}$ of metals in dust is required to explain the 
surface abundances of ${\lambda}$ Boo stars.  While more sophisticated treatments  are required to the 
accretion/diffusion models, our results at least show that metal infall at the required rate appears
to be occuring around ${\lambda}$ Boo  stars.          

In models for the first 100 Myr of the Solar System, Petit, Morbidelli \& Chambers (2001) have computed that approximately 50\% of the ${\sim}$5 M$_{\oplus}$ of planetesimals initially within the asteroid belt had their
orbits perturbed by Jupiter so they hit the Sun.  Murray et al. (2001) have
argued that such models can be generally extended to other main sequence
stars.  Assuming that this accretion of ${\sim}$5 ${\times}$ 10$^{12}$ g s$^{-1}$ of rocky material often occurs, the atmospheres of the A stars
should be substantially polluted.  We speculate that the main sequence stars
with inner holes in their dust distributions may also be systems with
asteroidal belts where gravitational perturbations by giant planets leads
to a significant supply of metals to an atmosphere where they would be otherwise
depleted.  An infall of large rocks or asteroids may lead to observable variations in the abundance of the elements in the star's atmosphere (Cowley 1977) which could be different
from the effects of dust inflows occuring around
the ${\lambda}$ Boo stars.

\section{CONCLUSIONS}
We  have obtained {\it Spitzer}  mid-infrared spectra of 19 main-sequence stars. Our data are most useful for constraining the 
evolution of the solid material in  regions between ${\sim}$10 
AU and ${\sim}$60 AU from the stars. We conclude the following: 
\begin{itemize}
\item{There are no strong spectral features; we suggest that the typical
diameter of a dust particle is greater than 10 ${\mu}$m.}
\item{The infrared excesses can be explained by dust arising from collisions
 between parent bodies whose  minimum total mass lies  between 0.004 
M$_{\oplus}$  and 0.06 M$_{\oplus}$.  These derived minimum masses are comparable to the
total mass of KBOs in the Solar System.} 
\item{We estimate dust production rates of ${\sim}$10$^{10}$ g s$^{-1}$ around   ${\lambda}$ Boo and HR 1570; selective accretion of this material may help explain their peculiar surface abundances.}
\item{There appear to be inner truncations 
in the dust clouds at  48 AU, 11 AU, 52 AU and 54 AU around  HR 333, HR 506,  HR 1082 and  HR 3927, respectively.}  
\end{itemize}

This work is based on observations made with the 
{\it Spitzer Space Telescope}, which is operated by the Jet 
Propulsion Laboratory, California Institute of Technology 
under NASA contract 1407. Support for this work was provided 
by NASA through Contract Number 1257184 issued by 
JPL/Caltech.  F.K.  is supported by the {\it Spitze}r Fellowship Program under award 011 808-001.  Work at Cornell and the University of Rochester has been supported by JPL contract 960803 to Cornell and Cornell subcontract 31419-5714 to the University of Rochester.  We are grateful to the entire
team of scientists and engineers that have made {\it Spitzer} and the IRS function so well. 

\newpage
\begin{center}
Table 1 --  Stars with Excesses in the IRS data
\end{center}
\begin{tabular}{lrrrrrrrrrrr}
\hline
\hline
Star & Sp.   & $D$ & $L_{*}$ & $t_{*}$ & $K_{2}$ &$L_{30}/L_{*}$ &$T_{ex}$ &$M_{PB}$&t(exp)& AOR$^{3}$ \\
    &  & (pc) & (L$_{\odot}$) &  (Gyr) & & (10$^{-4}$) & (K) & (M$_{\oplus}$) & (s)\\
\hline
HR 333 & A3V & 83 & 34&0.1&0.16&1 &97&0.03 &24 &0003553536\\
HR 506 & F9V  & 17 & 1.5&0.3$^{1}$&0.15&0.9&94 &0.004&24&0003553792\\ 
HR 1082 & A3IV/V  & 74& 12  & 0.1$^{1}$ & 0.15&2&72&0.02&24&0003555328\\
HR 1570 & A0V  & 37 & 15 & 0.1$^{2}$& 0.14 &0.4&88& 0.005&24& 0003555840 \\
HR 3927 & A0V  & 98 &44 &0.1&0.21&1&98&0.06& 24&0003558656\\
${\lambda}$ Boo &  A0V  & 30 & 15  & 0.2$^{2}$ & 0.21 & 0.4&114 &0.01 & 24&0003559936\\
\hline
\end{tabular}
\\
\\
$^{1}$
Zuckerman \& Song (2004); $^{2}$Heiter, Weiss \& Paunzen (2002).  In this Table,  $K_{2}$ (10$^{13}$ Jy Hz) is defined  in equation (1). $^{3}$ The {\it Spitzer} aor data file identification number.    
\newpage
\begin{figure}
\epsscale{1}
\plotone{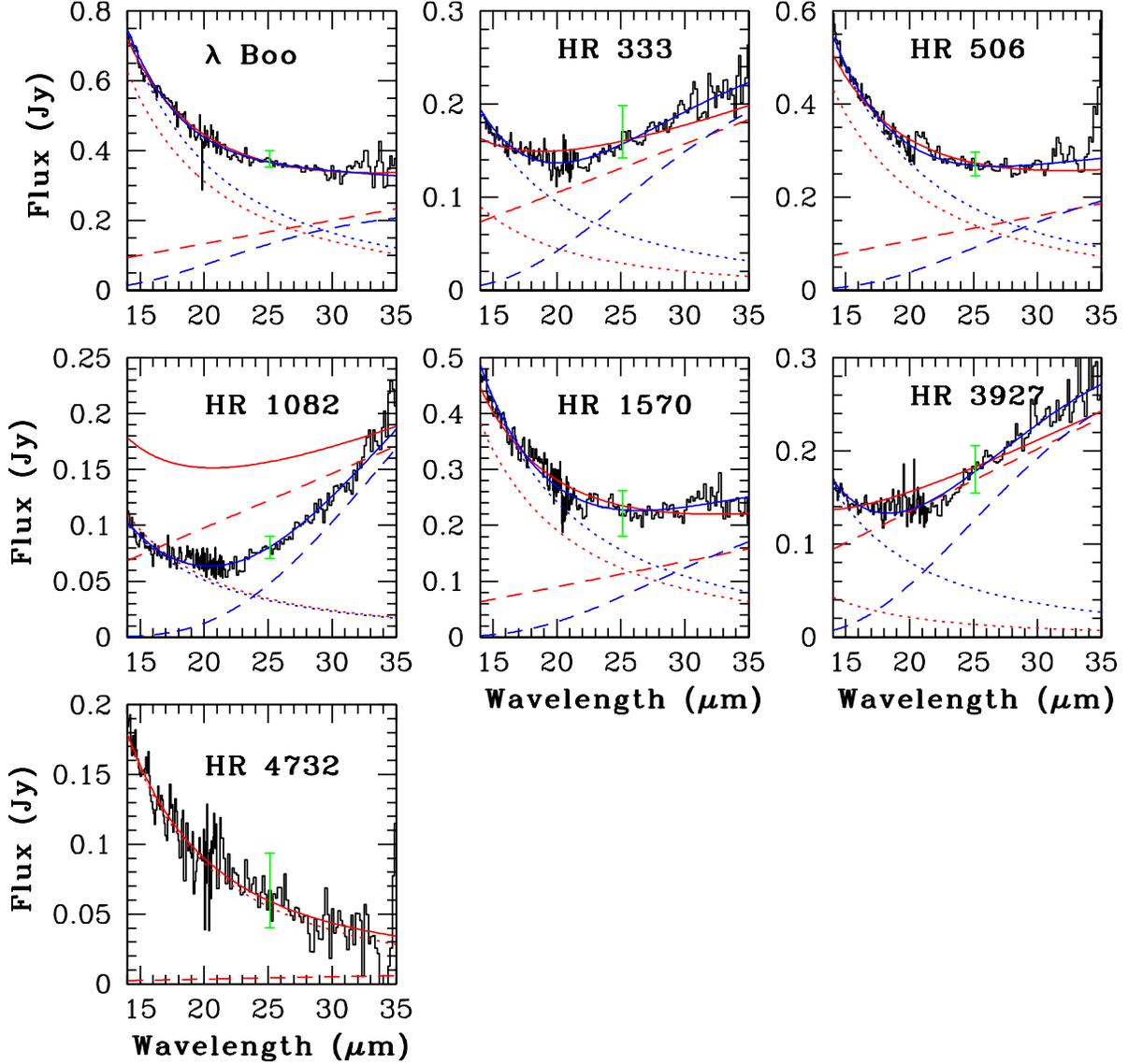}
\caption{Plots of $F_{\nu}$ vs. ${\lambda}$ for the 6 stars in Table 1 plus
HR 4732 which has, at most, a  modest infrared excess.  The  fits from equation (1) are shown in red while the fits from equation (2) are shown in blue.  We also display 3${\sigma}$ error bars at 25 ${\mu}$m in green. The sum of the two terms in each equation  is given by a solid line while dotted and dashed lines represent the photospheric fluxes and dust excesses, respectively.}
\end{figure}
\end{document}